# Three Dimensionality and Orbital Characters of the Fermi Surface in (Tl, Rb)$_y$Fe$_{2-x}$Se$_2$

Z.-H. Liu,[1] P. Richard,[2] N. Xu,[2] G. Xu,[2] Y. Li,[1] X.-C. Fang,[1] L.-L. Jia,[1] G.-F. Chen,[1] D.-M. Wang,[1]
J.-B. He,[1] T. Qian,[2] J.-P. Hu,[2,3] H. Ding,[2] and S.-C. Wang[1,*]

[1]*Department of Physics, Renmin University, Beijing, 100872, People's Republic of China*
[2]*Beijing National Laboratory for Condensed Matter Physics, and Institute of Physics, Chinese Academy of Sciences, Beijing 100190, People's Republic of China*
[3]*Department of Physics, Purdue University, West Lafayette, Indiana 47907, USA*

We report a comprehensive angle-resolved photoemission spectroscopy study of the tridimensional electronic bands in the recently discovered Fe selenide superconductor (Tl, Rb)$_y$Fe$_{2-x}$Se$_2$ ($T_c$ = 32 K). We determined the orbital characters and the $k_z$ dependence of the low energy electronic structure by tuning the polarization and the energy of the incident photons. We observed a small 3D electron Fermi surface pocket near the Brillouin zone center and a 2D like electron Fermi surface pocket near the zone boundary. The photon energy dependence, the polarization analysis and the local-density approximation calculations suggest a significant contribution from the Se $4p_z$ and Fe $3d_{xy}$ orbitals to the small electron pocket. We argue that the emergence of Se $4p_z$ states might be the cause of the different magnetic properties between Fe chalcogenides and Fe pnictides.

The recently discovered Fe chalcogenide superconductors $A_y$Fe$_{2-x}$Se$_2$ ($A$ = K, Rb, Cs, Tl, etc.) with maximum $T_c$ = 32 K [1,2] changed the landscape of Fe-based superconductivity. Unlike the other Fe-based superconductors, for which both hole-like and electron-like Fermi surfaces (FS) are observed simultaneously [3–9], these new materials are characterized by the absence of hole-like FS [10–13]. As with most of the other materials though, nearly isotropic superconducting (SC) gaps in the strong coupling regime are found on the observed FSs [11–13], suggesting a possible universal pairing mechanism for all Fe-based superconductors. Local antiferromagnetic (AFM) exchange interactions emerge as a serious candidate for the unification and the $A_y$Fe$_{2-x}$Se$_2$ system constitutes a good test to this approach. Indeed, neutron scattering experiments [14,15] suggested a magnetic ground state of the parent compound that differs from the collinear order commonly observed in the Fe pnictides [16,17]. In addition, large AFM exchange coupling constants have been determined experimentally in $A_y$Fe$_{2-x}$Se$_2$ as well as in FeTe [15,18] for the next ($J_2$) and next-next ($J_3$) nearest neighbors. Unfortunately, there is no experimental understanding of the origin of these enhanced constants that tune the magnetic ground state of these materials and may as well tune their SC properties.

Here, we report an angle-resolved photoemission spectroscopy (ARPES) study on SC $A_y$Fe$_{2-x}$Se$_2$ ($T_c$ = 32 K). We reveal an electron band with strong photon energy ($h\nu$) dependence, forming a 3D electron FS centered at $Z$ = (0, 0, $\pi$), whose intensity is enhanced with $s$ polarized light compared to $p$ polarized light. From local-density approximation (LDA) calculations and experimental observations, we conclude that this band has significant Fe $3d_{xy}$ and Se $4p_z$ orbital components. As a consequence, we propose that the Fe $3d_{xy}$ and Se $4p_z$ orbitals at $E_F$ enhance the superexchange constants and tune the magnetic structure of the system.

High quality single crystals with nominal composition (Tl, Rb)$_{0.8}$Fe$_{1.63}$Se$_2$ were synthesized by the flux method. The SC transition temperature $T_c$ = 32 K with a transition width of 2 K was measured with a physical property measurement system [2]. Samples were mounted in an inert gas protected glove box to prevent reaction with moisture. Samples with size smaller than $1 \times 1$ mm$^2$ were cleaved *in situ*, yielding a flat mirror like (001) surface. ARPES measurements were performed at Renmin University of China with a Scienta R3000 analyzer and at the Apple-PGM and PGM beam lines of the Synchrotron Radiation Center (Stoughton, WI), with Scienta 200U and Scienta R4000 analyzers, respectively. Normal state spectra were taken at $T$ = 40 K. During measurements, the pressure was maintained below $4 \times 10^{-11}$ Torr and no noticeable aging was observed during measurement cycles. The $E_F$ was referenced to a fresh Au polycrystalline film in good electrical contact to the sample.

The unfolded Brillouin zone (BZ) of (Tl, Rb)$_y$Fe$_{2-x}$Se$_2$ and high symmetry points are presented in Fig. 1(a) and the experimental setup is shown in Fig. 1(d). The band dispersion $E_B(k)$ along the $c$ axis can be measured by changing the photon energy ($h\nu$) with $k_z$ estimated using the formula [19,20]: $k_z = \frac{1}{\hbar}\sqrt{2m_e[(h\nu - \phi - E_B + V_0) - \hbar^2 k_\parallel^2}$, where $\phi$ is the work function and $V_0$ the inner potential of the sample. In (Tl, Rb)$_y$Fe$_{2-x}$Se$_2$, with the lattice constant $c$ = 14.56 Å [21], $\phi$ = 4.3 eV and empirically determined

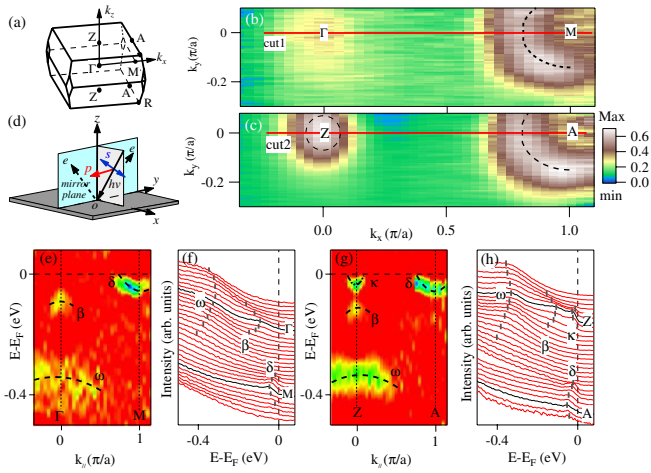
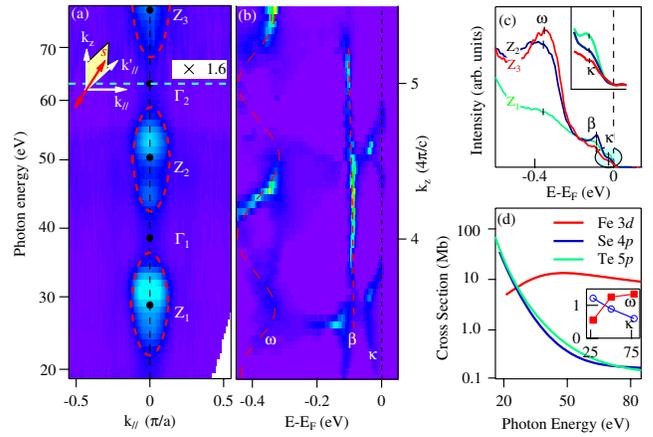

FIG. 1 (color online). (a) Schematic unfolded BZ of $(Tl, Rb)_y Fe_{2-x} Se_2$. (b), (c) Integrated intensity plots at $E_F$ ($-10$ meV, 10 meV) of $(Tl, Rb)_y Fe_{2-x} Se_2$ in the $Z$ ($h\nu = 28$ eV) and $\Gamma$ ($h\nu = 38$ eV) planes, respectively. Dashed lines are extracted FSs. (d) Sketch of the experimental polarization setup. The $s$ and $p$ polarizations are for electric fields perpendicular and parallel to the mirror plane (cyan), respectively. We note that our experimental setup for the $s$ geometry includes a polarization component along $z$. (e) Second derivative $[\partial^2 I(\omega, k)/\partial \omega^2]$ plots of band dispersion along $\Gamma$-$M$, indicated as cut1 in (b). (f) Corresponding energy-distribution curves (EDCs). Dashed lines are guides to the eye for the band dispersions. (g), (h) Same as (e) and (f) but along $Z$-$A$.

FIG. 2 (color online). (a) ARPES intensity plot in the $h\nu$-$k_\parallel$ plane of $(Tl, Rb)_y Fe_{2-x} Se_2$ ($\pm 10$ meV integration centered at $E_F$). The intensity above the cyan dashed line has been multiplied by 1.6. (b) Intensity plot of EDC curvature [25] at $k_\parallel = (0, 0)$ with different $h\nu$ values. The band dispersions of the $\omega$ and $\beta$ bands along $k_z$ are indicated by dashed lines serving as guides to the eye. (c) Comparison of EDCs of the $\kappa$ band close to $Z_1$, $Z_2$, and $Z_3$. The relative intensity of the $\kappa$ and $\omega$ bands vs. $h\nu$ are plotted in the inset of (d). (d) Calculated atomic photon-ionization cross sections for Fe $3d$, Se $4p$, and Te $5p$ orbitals [26]. The electric field is marked as red arrows in panel (a).

$V_0 = 15$ eV, we found that $h\nu = 18$, 38 and 63 eV are close to the $\Gamma$ point, while 28, 50 and 78 eV are close to the $Z$ point.

Figures 1(b) and 1(c) compare the normal state FSs obtained in the $\Gamma$ plane ($h\nu = 38$ eV) and in the $Z$ plane ($h\nu = 28$ eV), respectively. The FSs are in agreement with previous reports on the electronic structure of $A_y Fe_{2-x} Se_2$ [10–13]. We identify a FS near the BZ boundary ($M$, $A$), which is almost $k_z$ independent [22,23]. This observation is supported by the band dispersions, which are given in Figs. 1(e)–1(h) for the $\Gamma$-$M$ and $Z$-$A$ high symmetry lines, respectively [24]. Compared with the Fe pnictides, the hole-like bands near the 2D zone center are shifted down to about 100 meV below $E_F$ and thus do not contribute to the FS, as reported previously [10–13].

The main contrast between the $\Gamma$ and $Z$ plane lies in the existence of an additional $Z$-centered electron-like band, $\kappa$, which is observed in the $Z$ plane and fully absent in the $\Gamma$ plane. In the $Z$ plane, the $\kappa$ band has its minimum at 20–30 meV below $E_F$ and crosses $E_F$ at $k_\parallel \approx 0.1\pi/a$, yielding the small $Z$-centered FS pocket observed in Fig. 1(c). To investigate the detailed band dispersion along the $k_z$ direction, we did ARPES measurements with $h\nu$ ranging from 15 to 80 eV, which covers $k_z$ from $Z_1$ to $Z_3$. Figure 2 shows the $h\nu$-$k_\parallel$ FS intensity plot, with $k_\parallel$ oriented along the $Z$-$A$ line. We observe three FSs centered at 28

($Z_1$), 50 ($Z_2$) and 78 eV ($Z_3$) respectively, with the FSs lying halfway between $Z$ and $\Gamma$. We note that the total enclosed FS area of the $\kappa$ and $\delta$ bands is about 6.5% of the unfolded BZ (13% electron per Fe), confirming electron doping in this system.

The $h\nu$ variation of the $\kappa$ band is also well illustrated by the intensity plot of EDC curvature [25] along $(0, 0, k_z)$, as shown in Fig. 2(b). The $\kappa$ band has a clear dispersion along the $k_z$ ($h\nu$) direction, with its band minimum at $Z$ ($Z_1$ and $Z_2$) and across $E_F$ between $Z$ and $\Gamma$. Interestingly, a direct comparison of the EDCs at 28 ($Z_1$), 50 ($Z_2$), and 78 eV ($Z_3$), displayed in Fig. 2(c), reveals that the spectral intensity associated with the $\kappa$ band decreases with $h\nu$. Besides the $\kappa$ band, the $\omega$ band lying around 350–450 meV below $E_F$ also exhibits appreciable $h\nu$ variation, with a band maximum at $Z$ and a minimum at $\Gamma$. However, its spectral intensity evolves just the opposite way, with a stronger peak at $Z_3$ than at $Z_2$ and $Z_1$. This has an important consequence: although the strongly $k_z$ dispersive nature of these bands indicate that they are unlikely related to a surface state and that they carry significant components from orbitals extended along the $c$ axis, their respective dominant orbital or elemental characters must be different. The photoemission cross sections [26] of the Fe $3d$ and Se $4p$ states within the $h\nu = 20$–80 eV range are plotted in Fig. 2(d). In contrast to the spectral weight of the $\omega$ band, the $h\nu$ variation of the spectral weight of the $\kappa$ band, as shown in the inset of Fig. 2(d), cannot be described by the cross section calculated for Fe $3d$ states. Surprisingly, it shows much better agreement with the cross section

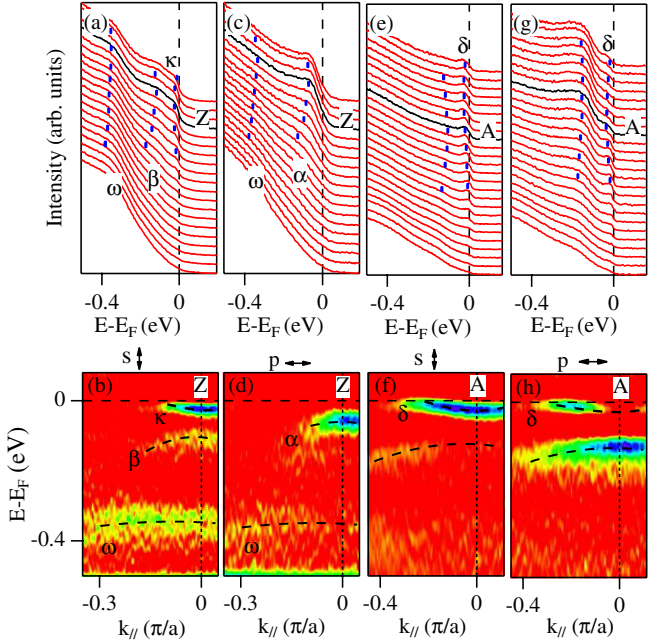

FIG. 3 (color online). Photon polarization dependence of the electronic structure of $(Tl, Rb)_y Fe_{2-x} Se_2$ in the $Z$ plane ($h\nu = 28$ eV). (a) The EDC intensity plot along the $Z$-$A$ direction, obtained with $s$ polarized incident photons. (b) The second derivative intensity plot along the same cut as in (a). (c), (d) Same as (a) and (b) but with $p$ polarized light. (e), (f) Same as (a) and (b) but with $s$ polarized light at $A$. (g), (h) Same as (a) and (b) but with $p$ polarized light at $A$. The electric field polarization is indicated as $s$, $p$, and corresponding arrows in all second derivative intensity plots.

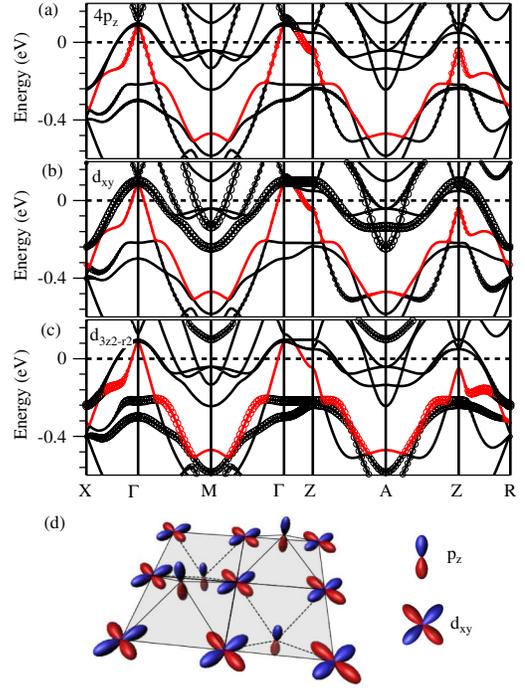

FIG. 4 (color online). LDA calculations and orbital projections (given by the symbol size) of $Tl_{0.8}Fe_2Se_2$ along high symmetry lines. (a) The band dispersion and projections onto the Se $4p_z$ orbital. (b), (c) Band dispersion and projections onto the Fe $3d_{xy}$ and Fe $3d_{3z^2-r^2}$ orbitals, respectively. The LDA calculations have been shifted by 190 meV to account for the electron doping in this sample and renormalized by a factor of 2.5 due to correlation effects [10]. (d) Sketch of the Fe $3d_{xy}$ and Se $4p_z$ orbitals.

calculated for Se $4p$ states, which monotonically decreases in the considered $h\nu$ range.

To refine our assumptions on the nature of the electronic states in the vicinity of $E_F$, we performed a photon polarization dependence study. Fig. 3 shows the bands in the $Z$ plane with $s$ and $p$ polarized incident photons, where $s$ and $p$ refer to an electric field perpendicular and parallel to the photoemission plane defined by the normal of the sample surface and the photoemitted electrons, respectively. (we note that our $s$ experimental geometry contains a component of the light polarization along the $z$ axis). Along the $Z(\Gamma)$-$A(M)$ direction, the $\alpha$ band is visible using $p$ polarization but suppressed with $s$ polarization. The $\beta$ band behaves oppositely. By analogy with the $Ba_{0.6}K_{0.4}Fe_2As_2$ and $FeTe_{0.55}Se_{0.45}$ compounds [27], this suggest that the the $\alpha$ and $\beta$ bands are likely related to the Fe $3d_{xy}$ and the even combination of the Fe $3d_{xz}$ and Fe $3d_{yz}$ orbitals, respectively. The electron-like band $\delta$ at the zone boundary $A$, with a minimum at $\sim 50$ meV, is seen in both configurations, suggesting a complicated orbital character that deserves further detailed study. While the $\kappa$ band at $Z_1$ is clearly observed with $s$ polarized light, as shown in Figs. 3(a) and 3(b), it is barely detectable with pure $p$ polarization [see Figs. 3(c) and 3(d)]. The photon polarization dependence of the $\kappa$ band is consistent with either a wave function having odd symmetry with respect to the measurement plane or with a $z$-oriented wave function.

To clarify the situation, we present in Fig. 4 our LDA calculations for $TlFe_2Se_2$ using the experimental structure parameters [28], along with projections for the Se $4p_z$, Fe $3d_{xy}$ and Fe $3d_{3z^2-r^2}$ orbitals. As with previous LDA calculations [29–32], we notice discrepancies with ARPES observations [10–13]. The main one is a downshift of the hole bands near the zone center accompanied by an upshift of the electron bands near the zone boundary, which has been explained in terms of correlation effects and the evidence of interband coupling [10,33]. Nevertheless, LDA captures several important features of the ARPES observations [10]. LDA calculations indicate that the energy of the $\kappa$ band is sensitive to the height of the Se atoms and shifts down below $E_F$ in $(Tl, Rb)_y Fe_{2-x} Se_2$ [29]. Interestingly, the LDA orbital projections suggest that in addition to the Se $4p_z$ orbital character, the $\kappa$ band also carries a non-negligible Fe $3d_{xy}$ component, thus explaining its sensitivity to odd configurations of light polarization. The calculations also suggest that the $k_z$ dispersive $\omega$ band at higher binding energy origins mainly from the Fe $3d_{3z^2-r^2}$ orbital.

It has been recently proposed that local AFM exchange fluctuations [34], which are characterized by exchange coupling constants, mainly determine the momentum distribution of the SC gap in high-temperature superconductors such as cuprates and Fe-based superconductors [35]. The $A_y\text{Fe}_{2-x}\text{Se}_2$ system has a complicated magnetic structure that differs from that of the Fe pnictides [14,15,36]. A recent neutron scattering study reported a ferromagnetic nearest neighbor exchange coupling ($J_1$), appreciable AFM next nearest ($J_2$) and next-next nearest neighbor couplings ($J_3$), as well as an AFM interlayer coupling ($J_c$) in $\text{Rb}_y\text{Fe}_{2-x}\text{Se}_2$ [15]. It is widely believed that the magnetic super-exchange interactions in Fe-based superconductors are at least partially As/Se bridged [37]. Our experimental observation of Se $4p_z$ states hybridized with Fe $3d$ orbitals at $E_F$ in $(\text{Tl}, \text{Rb})_y\text{Fe}_{2-x}\text{Se}_2$ provides a natural explanation to the existence of a large $J_3$ AFM exchange coupling. An intuitive picture can be argued as follows. Let's consider a Fe atom. A local Wannier state can be created because the Fe $3d_{xy}$ orbital strongly hybridizes with the $4p_z$ orbital of its four Se atom neighbors, as illustrated in Fig. 4(d). This Wannier state can link the two neighboring Fe atoms on each side through a standard super-exchange interaction, thus resulting in the $J_3$ AFM coupling. In this picture, the $J_3$ is very sensitive to the hybridization due to the size of the $p_z$ orbital and the distance of the Se/Te atoms from the Fe-layers. Interestingly, a possible hybridization between Fe $3d$ and $p_z$ states has also been noted in the cousin $\text{FeSe}_{0.45}\text{Te}_{0.55}$ system [6], which is also characterized by a large $J_3$ AFM coupling.

To summarize, we performed ARPES measurements on SC $(\text{Tl}, \text{Rb})_y\text{Fe}_{2-x}\text{Se}_2$ with various $h\nu$ values and different photon polarizations. We observed a small $Z$-centered electron-like FS pocket. LDA calculations and experimental data suggest that it originates mainly from a mixture of the Se $4p_z$ and Fe $3d_{xy}$ orbitals. These orbitals at $E_F$ enhance the exchange coupling between Fe ions bridged by Se and may play an important role in the superconductivity of the $A_y\text{Fe}_{2-x}\text{Se}_2$ family.

This work was supported by the grants from the Chinese Academy of Sciences (2010Y1JB6), the Ministry of Science and Technology of China (2010CB923000, 2012CB821400 and 2011CBA001001), the National Science Foundation of China (10774189, 10974175, 11004232, 11190024 and 11050110422) and the Research Funds of Renmin University of China (12XNH093). This work is based in part upon research conducted at the Synchrotron Radiation Center, which is primarily funded by the University of Wisconsin-Madison with supplemental support from facility users and the University of Wisconsin-Milwaukee.